\documentclass[iop,revtex4,numberedappendix,twocolappendix]{emulateapj}
\usepackage{apjfonts}
\usepackage{epsfig,graphicx}
\usepackage{amssymb,amsmath}
\usepackage{latexsym}
\usepackage{multirow}
\usepackage{natbib}
\usepackage{lscape}
\usepackage[maxfloats=36]{morefloats}
\usepackage{longtable}
\usepackage{enumitem}
\usepackage{afterpage}  
\usepackage{comment}

\shorttitle{The first field BH-LMXB system identified in quiescence}
\shortauthors{Tetarenko, B.E. et al.}

\submitted{Submitted Version: \today}

\begin{document}
\title{The first low-mass black hole X-ray binary identified in quiescence outside of a globular cluster}

\author{B.E. Tetarenko\altaffilmark{1}, A. Bahramian\altaffilmark{1}, R.M. Arnason\altaffilmark{1,2}, J.C.A. Miller-Jones\altaffilmark{3}, S. Repetto\altaffilmark{4}, C.O. Heinke\altaffilmark{1}, T.J. Maccarone\altaffilmark{5}, L. Chomiuk\altaffilmark{6}, G.R. Sivakoff\altaffilmark{1}, J. Strader\altaffilmark{6}, F. Kirsten\altaffilmark{3}, and W. Vlemmings\altaffilmark{7}}
\affil{$^1$Department of Physics, University of Alberta, CCIS 4-181, Edmonton, AB T6G 2E1, Canada}
\affil{$^2$Department of Physics and Astronomy, University of Western Ontario, 1151 Richmond Street, London, ON N6A 3K7, Canada}
\affil{$^3$International Centre for Radio Astronomy Research (ICRAR), Curtin University, GPO Box U1987, Perth, WA 6845, Australia}
\affil{$^4$Physics Department, Technion - Israel Institute of Technology, Haifa, Israel 32000}
\affil{$^5$Department of Physics, Texas Tech University, Box 41051, Lubbock, TX 79409-1051, USA}
\affil{$^6$Department of Physics and Astronomy, Michigan State University, East Lansing, MI 48824, USA}
\affil{$^7$Department of Earth and Space Sciences, Chalmers University of Technology, Onsala Space Observatory, SE-439 92 Onsala, Sweden}
 
\email{btetaren@ualberta.ca}   

\begin{abstract}
The observed relation between the X-ray and radio properties of low-luminosity accreting black holes has enabled the identification of multiple candidate black hole X-ray binaries (BHXBs) in globular clusters. Here we report an identification of the radio source VLA J213002.08+120904 (aka M15 S2), recently reported in \citealt{kirsten2014}, as a BHXB candidate. They showed that the parallax of this flat-spectrum variable radio source indicates a 2.2$^{+0.5}_{-0.3}$ kpc distance, which identifies it as lying in the foreground of the globular cluster M15.  We determine the radio characteristics of this source, and place a deep limit on the X-ray luminosity of $\sim4\times10^{29}$ erg s$^{-1}$. Furthermore, 
we astrometrically identify a faint red stellar counterpart in archival Hubble images, with colors consistent with a foreground star; at 2.2 kpc its inferred mass is 0.1-0.2 $M_{\odot}$.
We rule out 
that this object is a pulsar, neutron star X-ray binary, cataclysmic variable, or planetary nebula, concluding that VLA J213002.08+120904 is the first accreting black hole X-ray binary candidate discovered in quiescence outside a globular cluster. Given the relatively small area over which parallax studies of radio sources have been performed, this discovery suggests a much larger population of quiescent BHXBs in our Galaxy, $2.6\times10^4-1.7\times10^8$ BHXBs at $3\sigma$ confidence, than has been previously estimated ($\sim10^2-10^4$) through population synthesis.

\end{abstract}

\maketitle
\section{Introduction}\label{s:intro}
Black hole X-ray binaries (BHXBs) are interacting binary systems where X-rays are produced by material accreting from a secondary companion star onto a black hole (BH) primary.
Due to angular momentum in the system, accreted material does not flow directly onto the compact object; rather, it forms a differentially rotating disk around the BH known as an accretion disk \citep{ss73}.
While some material accretes onto the BH, a portion of this inward falling material may also be removed from the system via an outflow in the form of a relativistic plasma jet or an accretion disk wind  \citep{blandford79,whiteholt82}.
For major reviews of BHXBs see \citet{chen97,mr06} and \citet{d07}.

Currently, the known Galactic BHXB population is made up of 19 dynamically confirmed BHs, and $60$ black hole candidates (BHCs), 56 of which are located in the field \citep{tetarenkob2015}. The remaining four are found in globular clusters (GCs; \citealt{strader2012,chomiuk2013,millerjones2015b}). The vast majority of these Galactic BHXBs are low-mass X-ray binaries (LMXBs), where mass transfer occurs via Roche lobe overflow of a secondary companion with a mass $M_2 \lesssim 3 \,  M_{\odot}$ and spectral type A or later. In addition, most of these systems are transient, cycling between periods of quiescence and outburst. This behaviour is associated with changing geometries of mass inflow and outflow (see e.g., \citealt{mr06}).

While the majority of the known population has been discovered during outburst through bright X-ray emission, typically peaking between $\sim{ \rm 10^{36}-10^{40} \, erg \, s^{-1}}$ (e.g., \citealt{chen97,tetarenkob2015}), the combination of more sensitive X-ray and radio telescopes have enabled detection and identification of a handful of candidate quiescent BH-LMXB systems in three Galactic GCs. Currently, no BH-LMXBs known to exist in the field have been first identified outside of outburst and no X-ray outbursts have been clearly identified from candidate BH-LMXBs in Galactic GCs. This is not surprising as it is far more difficult to identify quiescent BHs in the field, compared to those in GCs. Without established distances to field sources, simply having the ratio of X-ray to radio flux can not strongly rule out background galaxies.

Typical estimates for the total number of BH-LMXBs in the Galaxy have been computed both empirically using large-scale observational surveys (e.g., \citealt{romani1998,kalogera1999,corralsantana2015}), and theoretically from population synthesis codes (e.g., \citealt{romani1992,romani1994,portegieszwart1997,kalogera1998,pfahl2003,yungelson2006,kiel2006}). These estimates span a wide range, on the order of $\sim10^2-10^4$.
This large uncertainty is due to poorly constrained key characteristics that describe the Galactic binary population including, the BH natal kick distribution, initial stellar mass function, the binary fraction, distribution of binary periods and mass ratios, and the physics of common envelope and binary stellar evolution (e.g., \citealt{portegieszwart1996,belczynski2002,pfahl2003,ivanova2013}). Detecting and identifying more of these systems is crucial to observationally constraining these uncertain parameters.

Accreting BHs are known to (i) emit compact radio jets in both the hard and quiescent spectral states (e.g., \citealt{gallo2004,rm06,d07} for a description of X-ray spectral states) that produce a flat, to slightly inverted spectrum ($\alpha\sim0.0$ to $0.7$ where $F_{\nu}\propto \nu^{\alpha}$; \citealt{blandford79,fbg04,fe09,russ12,rus13}), (ii) be more radio bright (i.e., have much higher radio luminosities at a given X-ray luminosity) than neutron stars \citep{fenderhendry2000,migliari2006}, and (iii) exhibit a strong correlation 
between X-ray and radio emission in the hard and quiescent spectral states (e.g., \citealt{hann98,corb0,corb3}). 
As such, it is possible to distinguish between an accreting BH and an analogous neutron star or white dwarf system with a combination of X-ray and radio observations of a source. Specifically, we can use a technique, originally suggested by \cite{maccarone2007}, that combines the now established ``universal'' correlation between X-ray and radio luminosity in BHXBs \citep{gallo2003} with spectral shape and position of a source in the radio/X-ray ($L_R/L_X$) plane.

Here we report on the new BH candidate VLA J213002.08+120904 (hereafter known as VLA J2130+12), the first field BH-LMXB system identified by quiescent emission. VLA J2130+12 is located at a right ascension and declination of 21h 30m 02.086s, and $12^{\circ}$09\arcmin 04.220\arcsec, respectively.
Through a seven epoch VLBI campaign, \citealt{kirsten2014} (hereafter K14) showed that VLA J2130+12, aka M15 S2 \citep{knapp1996},
is not associated with M15, and argued that it is likely a field LMXB.
K14 derive a parallax of $\pi=0.45\pm0.08$ mas, indicating a distance of $2.2^{+0.5}_{-0.3}$ kpc. Comparing this to M15's distance of $10.3\pm0.4$ kpc \citep{vandenbosch2006} proves that VLA J2130+12 is not related to M15.
They find a proper motion of VLA J2130+12 of $(\mu_{\alpha},\mu_{\delta})=(-0.07\pm0.13,-1.26\pm0.29) \, {\rm mas \, yr^{-1}}$, which is also inconsistent with the proper motion found by K14 for M15 members, average $(\mu_{\alpha},\mu_{\delta})=(0.58,4.05)  \, {\rm mas \, yr^{-1}}$. K14 found radio flux variations in VLAJ2130+12 by a factor of 6 on timescales of a few months. Despite the known variability, they combined their 1.6 GHz data with higher frequency data reported in \citealt{knapp1996} ($230\pm40$ $\mu$Jy at 8.4 GHz) to suggest that the source has a flat radio spectrum. This would be indicative of an XRB.

In Section 2 we describe the observations and reduction of our Chandra and VLA data, as well as the archival HST data that we make use of. In Section 3 we present our observational results and argue for the presence of a stellar-mass BH in VLA J2130+12. In Section 4 we discuss the nature and system parameters of VLA J2130+12, refute alternative explanations, and provide an empirical estimate for the total of the number of BH-LMXBs in the Galaxy. Lastly, we provide a summary of our results in Section 5.

\begin{center}
\begin{figure}[h]%
\plotone{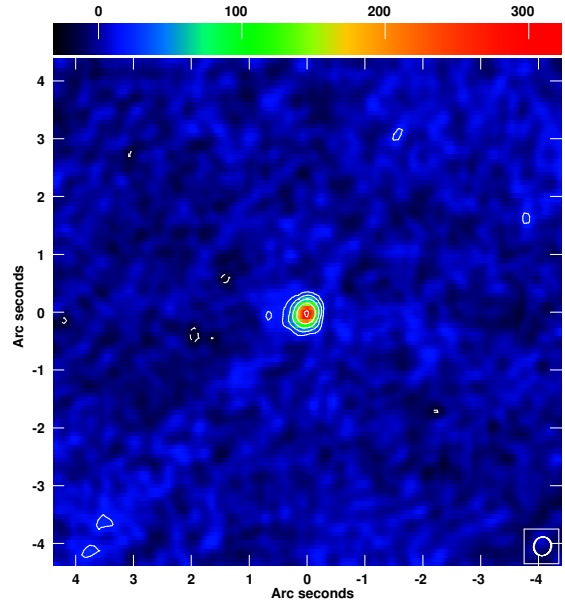}
  \caption{Radio image of VLA J2130+12 made from the data taken on 2011 August 21.  The image was made with contour levels of $\pm 2^n$ times the lowest contour (with $n=1,2,3$ etc.), where the lowest contour level is three times the rms noise of 6.2 $\mu$Jy/beam. The colour bar is in units of $\mu$Jy/beam.  The source position agrees exactly with the VLBI position of K14.}%
    \label{fig:radioimage}%
\end{figure}
\end{center}

\section{Observations and Data Reduction}

\subsection{VLA Data}

In 2011 May, MAXI detected an X-ray flare in M15 \citep{morii2011}.  A Swift/XRT observation subsequently localized the event to the cluster core \citep{heinke2011}, leading us to make Target-of-Opportunity observations with the Karl G. Janksy Very Large Array (VLA), under program code 11A-269. \citet{straderjay2012} reported an analysis of the cluster core using these data, placing a limit on radio emission from a possible intermediate-mass black hole at the cluster center. We observed M15 over five epochs, for a total of 10\,hr of time (Table~\ref{table:vladata}).  The array was in its extended hybrid BnA-configuration on 2011 May 22, 26, and 30 and in its most-extended A-configuration for the final two observations on 2011 August 21 and 22.  We observed in two 1024-MHz basebands, centred at frequencies of 5.0 and 7.45\,GHz.  Each baseband comprised eight spectral windows, each of which consisted of 64 spectral channels of width 2\,MHz.  We reduced the data using standard procedures within the Common Astronomy Software Application \citep[CASA;][]{mcmullin2007}.  We used 3C\,48 to set the amplitude scale (according to the Perley-Butler 2010 coefficients within the CASA task {\sc setjy}) and to perform both the instrumental delay calibration and the bandpass calibration.  We used the nearby, compact calibrator J2139+1423 to 
determine the complex gain solutions that were interpolated onto the target field.  The calibrated data were then imaged with natural weighting, for maximum sensitivity.  The initial calibration was good, and with only a few mJy of emission (dominated by a bright source at the edge of the field) we did not attempt self-calibration.

While the main science target was the outburst of M15 X-2 \citep{millerjones2011a}, the source VLA J2130+12 was also within the field, at the 94 and 87\% response points of the primary beam at 5.0 and 7.45\,GHz, respectively.  We corrected the images for the effect of the primary beam response, and then measured the flux density of VLA J2130+12 (see Table \ref{table:vladata}) by fitting it with a point source model in the image plane, using the CASA task {\sc imfit}. See Figure \ref{fig:radioimage} for the radio image of VLA J2130+12. 

In addition, we searched the NRAO archive for past data
\afterpage{
\renewcommand{\thefootnote}{\alph{footnote}}
\renewcommand\tabcolsep{5pt}
\begin{longtable*}{lcccccr}

\caption{New 2011 VLA observations of M15}  \\

\hline \hline \\[-2ex]
   \multicolumn{1}{l}{Date} &
   \multicolumn{1}{c}{MJD } &
     \multicolumn{1}{c}{Duration}  &
        \multicolumn{1}{c}{Array} &
     \multicolumn{1}{c}{5 GHz} &
       \multicolumn{1}{c}{7 GHz}  &
         \multicolumn{1}{c}{Spectral}     \\
        && \multicolumn{1}{c}{(min)}& \multicolumn{1}{c}{Configuration}&\multicolumn{1}{c}{Flux Density}&\multicolumn{1}{c}{Flux Density}&\multicolumn{1}{c}{Index}\\
         && & &\multicolumn{1}{c}{($\mu$Jy/bm)}  &\multicolumn{1}{c}{($\mu$Jy/bm)}  &     \\[0.5ex] \hline
   \\[-1.8ex]
\endfirsthead

  \\[-1.8ex] \hline \\[-1.0ex] 
\endlastfoot
2011-05-22 & $55703.53\pm0.01$ & 60 & BnA&$508.1\pm9.1$ & $454.4 \pm11.9$& $-0.33\pm0.16$ \\[0.1cm]
2011-05-26 & $55707.50\pm0.02$ & 60 & BnA&$393.1\pm10.0$ & $420.6\pm12.0$& $0.17\pm0.11$\\[0.1cm]
2011-05-30 & $55711.51\pm0.06$ & 180 & BnA&$324.8\pm6.0$ & $337.9\pm8.5$& $0.10\pm0.08$\\[0.1cm]
2011-08-21 & $55794.34\pm0.05$ & 150 & A&$302.0\pm12.5$ & $330.5\pm10.7$& $0.23\pm0.15$\\[0.1cm]
2011-08-22 & $55795.44\pm0.05$ & 150 & A&$312.2\pm8.4$ & $334.4\pm7.9$& $0.17\pm0.10$\\[-2.5ex]
\label{table:vladata}
\end{longtable*}
     \renewcommand{\thefootnote}{\arabic{footnote}}
     \renewcommand\tabcolsep{3pt}
}

\afterpage{
\renewcommand{\thefootnote}{\alph{footnote}}
\renewcommand\tabcolsep{3pt}
\begin{longtable*}{lccccccc}

\caption{Archival VLA observations of M15}  \\

\hline \hline \\[-2ex]
   \multicolumn{1}{l}{Program ID} &
   \multicolumn{1}{c}{Date} &
   \multicolumn{1}{c}{MJD } &
     \multicolumn{1}{c}{Duration}  &
        \multicolumn{1}{c}{Array} &
             \multicolumn{1}{c}{1.49 GHz} &
       \multicolumn{1}{c}{4.86 GHz}  &
         \multicolumn{1}{c}{8.44 GHz}     \\
      &  && \multicolumn{1}{c}{(min)}& \multicolumn{1}{c}{Configuration}&\multicolumn{1}{c}{Flux Density}&\multicolumn{1}{c}{Flux Density}&\multicolumn{1}{c}{Flux Density}\\
           && & & &\multicolumn{1}{c}{($\mu$Jy/bm)}  &\multicolumn{1}{c}{($\mu$Jy/bm)} &\multicolumn{1}{c}{($\mu$Jy/bm)}     \\[0.5ex] \hline
   \\[-1.8ex]
\endfirsthead

  \\[-1.8ex] \hline \\[-1.0ex] 
        \multicolumn{8}{p{0.7\columnwidth}}{$^a$ AJ186 and AK244 data have been concatenated together, including observations on 1990 April 9, 10, 12, and 17 (47990.6, 47991.6, 47993.7, and 47998.6).}\\
                \multicolumn{8}{p{0.7\columnwidth}}{$^b$ \citealt{knapp1996}.}\\
\endlastfoot
AM264&1989-02-19 & $47576.6$ & 60 & BnA&-&$217.2 \pm36.7$&-\\[0.1cm]
AJ186/AK244$^a$&1990-04-09 & $47990.6$ &300  & A&$135.3 \pm22.8$&-&-\\[0.1cm]
AM308 &1990-07-13 & $48085.3$ & 96 & BnA&-&-&$210.6 \pm18.7$\\[0.1cm]
AB0578$^b$&1991-02-16 & $48303.5$ & 90 & CnD&-&-&$230\pm40$\\[0.1cm]
AJ211&1991-08-08 & $48476.6$ & 90 & A&-&-&$158.7 \pm18.9$\\[0.1cm]
AM342&1991-12-21 & $48611.9$ &42 & BnA&-&$345.1 \pm35.0$&-\\[0.1cm]
AB1131&2004-10-14 & $53292.9$ &324 & A&-&-&$221.9 \pm8.4$\\[-2.5ex]
\label{table:vladataarc}
\end{longtable*}
     \renewcommand{\thefootnote}{\arabic{footnote}}
     \renewcommand\tabcolsep{3pt}
}

\noindent on M15 that can constrain the variability of M15 S2, and found a number of useful observations obtained with the historical 
(pre-upgrade) VLA, spanning 1989--2004 (see Table~\ref{table:vladataarc}). All observations, with the exception of the 1990 April L-band observations, were obtained in the standard continuum mode of the historical VLA, with 2 intermediate frequency pairs, each of 50 MHz bandwidth sampled in a single channel, and in full polarization. The 1990 April observations were obtained with 2 IFs, each of 50 MHz bandwidth sampled with 7 channels (to minimize bandwidth smearing) and only one (right-hand) polarization. All data were edited and calibrated using standard routines in AIPS, and imaged using \verb|IMAGR| in AIPS with a Briggs weighting scheme (robust=1). Flux densities were measured using JMFIT in AIPS.

\subsection{Chandra Data}
\begin{center}
\begin{figure*}%
\plotone{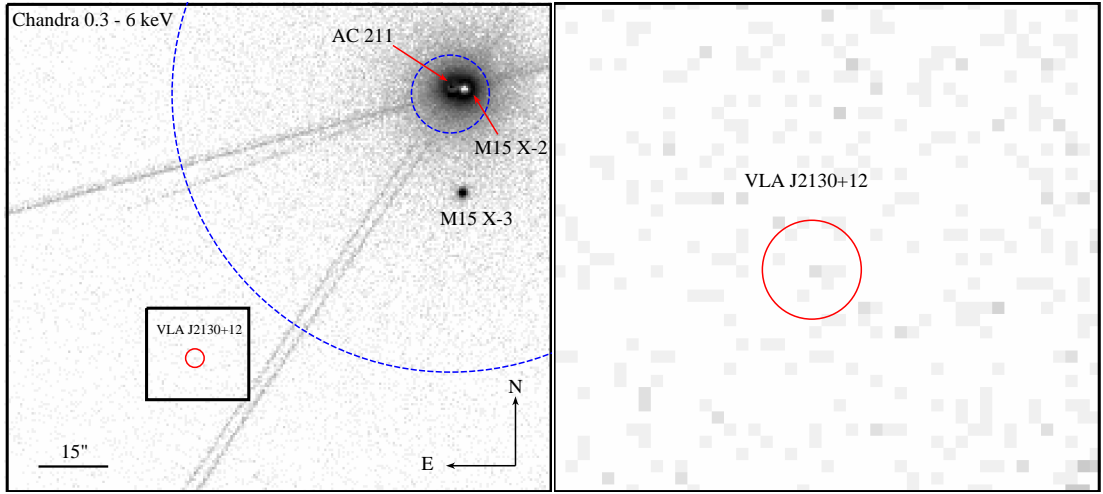}
  \caption{The left panel shows the merged (102 ks) Chandra/ACIS image in the 0.3--6 keV band. The red circle marks a 1\arcsec \,region around the location of S2 reported by K14. The locations of known sources in the field are also marked. AC211 and M15 X-2 are LMXBs \citep{white2001}, and M15 X-3 is an X-ray transient \citep{heinke2009}. The dashed blue contours mark the core and half-light radii of M15 (1 arcmin; \citealt{harris1996}). The right panel shows a zoomed in view of the black box region around VLA J2130+12 marked in the left panel. Note that the streaks in this image are artifacts caused by the large brightness of M15 X-2 and AC211.}%
    \label{fig:chandraimage}%
\end{figure*}
\end{center}

We analyzed all available non-grating Chandra/ACIS observations of M15 (Table~\ref{table:chandraobs}). 
All these observations were performed in faint mode and with the core of M15 on-axis. All observations were reduced and reprocessed using standard tools provided in CIAO 4.6 \citep{fruscione2006}. After reduction, all observations were combined using the task \textit{reproject\_obs} to obtain a total exposure of 102 ks (see Figure \ref{fig:chandraimage}). Note that we also checked all other Chandra data (HRC and grating mode observations) for detections of VLA J2130+12, finding no evidence for increased activity. We performed all analysis in the 0.3--6.0 keV band.

\subsection{HST Data}

We searched for the optical counterpart of VLA J2130+12 in data available from the Hubble Legacy Archive and ACS catalog \citep{sarajedini2007}. HST observations used in this work\footnote{Note that there are also WFPC2 observations of M15 in the F450W filter that covers the vicinity of VLA J2130+12. However, as WFPC2 has larger pixels, causing blending in crowded regions, we do not make use of these observations.} are tabulated in Table \ref{table:hstobs}. We corrected astrometry (e.g., fitted for rotation, scale, and shifts) in HST images in two steps. First, we used the 2MASS catalog and corrected astrometry of ACS F814W and F606W images using 
matching coordinates of 40 relatively bright sources detected in both 2MASS and ACS images. We chose only 2MASS sources whose ACS counterparts indicated that the 2MASS reported coordinates were not affected by blending of unresolved multiple bright sources. The 2MASS catalog absolute astrometry has an 1$\sigma$ uncertainty of $\sim$ 0.12 arcseconds\footnote{See 2MASS All-Sky Data Release User guide, Sec.~2.f, http://www.ipac.caltech.edu/2mass/releases/allsky/doc/sec2\_2.html and \citealt{skrutskie2006}.}. To increase the astrometric accuracy of the HST images further, in the second step, we used radio coordinates of two distinct sources detected in both VLA radio and ACS images, AC 211 and M15 S1. These sources are clearly detected in radio as reported by K14. AC 211 is clearly detected in all HST filters (identified through comparison with \citealt{white2001}). K14 indicate that S1 is a background AGN. In ACS images this source has a clear counterpart with a slightly-extended point spread function consistent with a background galaxy. After this (x,y) shift, our uncertainty in position in the HST images is reduced to 0.09 arcseconds (3$\sigma$). We applied astrometry corrections to the WFC3 images by matching sources detected in ACS and WFC3 filters.

\begin{center}
\begin{figure*}%
\plotone{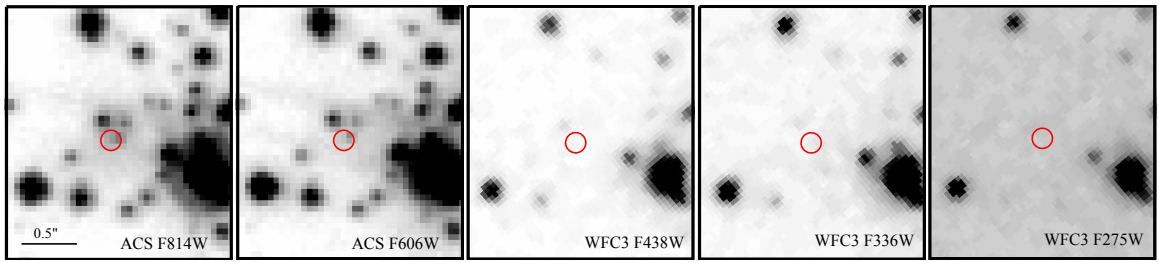}
  \caption{HST images of VLA J2130+12. The red circle indicates the radio position, as reported by K14, with an uncertainty of 0.09\arcsec. Individual images are labeled by detector and band. Note that the identified optical counterpart is detected in the ACS/WFC images but not clearly detected in the single-epoch WFC3/UVIS images (2011-10-22 data shown). }%
    \label{fig:hstimages}%
\end{figure*}
\end{center}

We find one star visible in the  ACS/WFC images (F814W and F606W filters), 0.08\arcsec \,away from the radio position of VLA J2130+12 reported in K14. Based on values reported in  the ACS catalog, the identified optical counterpart (See Figure \ref{fig:hstimages}) has an apparent (Vega) magnitude of $24.87\pm0.24$ in the V-band equivalent filter F606W and $23.09\pm0.12$ in I (F814W). In addition to the ACS survey images containing F814W and F606W filters, the vicinity of VLA J2130+12 was also

 \noindent observed with WFC3/UVIS in F438W, F336W and F275W (Table~\ref{table:hstobs}). However, the counterpart is not clearly detected in  single-epoch observations at any of these bands (See Figure \ref{fig:hstimages}).

\afterpage{
\renewcommand{\thefootnote}{\alph{footnote}}
\renewcommand\tabcolsep{5pt}
\begin{longtable}{lcccr}

\caption{Chandra/ACIS Observations of M15}  \\

\hline \hline \\[-2ex]
   \multicolumn{1}{c}{obsID} &
      \multicolumn{1}{c}{Instrument} &
   \multicolumn{1}{c}{Date} &
      \multicolumn{1}{c}{MJD} &
     \multicolumn{1}{r}{Exposure (ks)}  \\[0.5ex] \hline
   \\[-1.8ex]
\endfirsthead

  \\[-1.8ex] \hline \\[-1.0ex] 
\endlastfoot
0675&ACIS & 2000-08-24&51780 & 19.8 \\[0.1cm] 
1903&HRC-I & 2001-07-13&52103 & 9.1 \\[0.1cm] 
2412&HRC-I  & 2001-08-03& 52124& 8.82 \\[0.1cm] 
2413&HRC-I  & 2001-08-22&52143& 10.79 \\[0.1cm] 
4572&ACIS & 2004-04-17&53112 & 59.17 \\[0.1cm] 
9584&HRC-I  & 2007-09-05& 54348& 2.15 \\[0.1cm] 
11029&ACIS & 2009-08-26&55069 & 34.18 \\[0.1cm] 
11886 &ACIS& 2009-08-28&55071 & 13.62 \\[0.1cm] 
11030 &ACIS& 2009-09-23&55097 & 49.22 \\[0.1cm] 
13420&HRC-I  & 2011-05-30& 55711& 1.45 \\[0.1cm] 
13710&ACIS& 2012-09-18&56188 & 4.88 \\[0.1cm]
14618&HRC-S  & 2013-10-07&56572 & 15.1 \\[-2.5ex]
\label{table:chandraobs}
\end{longtable}
     \renewcommand{\thefootnote}{\arabic{footnote}}
     \renewcommand\tabcolsep{3pt}
}
\afterpage{
\renewcommand{\thefootnote}{\alph{footnote}}
\renewcommand\tabcolsep{5pt}
\begin{longtable}{lcccr}

\caption{Archival HST Observations of M15 used in this work}  \\

\hline \hline \\[-2ex]
   \multicolumn{1}{l}{Proposal ID} &
      \multicolumn{1}{c}{Date} &
   \multicolumn{1}{c}{Detector} &
     \multicolumn{1}{c}{Band}  &
        \multicolumn{1}{r}{Exposure (s)} \\[0.5ex] \hline
   \\[-1.8ex]
\endfirsthead

  \\[-1.8ex] \hline \\[-1.0ex] 
\endlastfoot
10775 &2006-05-02& ACS/WFC & F814W&615 \\[0.1cm] 
 &  && F606W&535 \\[0.1cm] 
12605&2011-10-16 & WFC3/UVIS & F438W&130 \\[0.1cm] 
& & & F336W&700 \\[0.1cm] 
&&  & F275W&1315 \\[0.1cm] 
&2011-10-22 & WFC3/UVIS & F438W&130 \\[0.1cm] 
& & & F336W&700 \\[0.1cm] 
& & & F275W&1315\\[-2.5ex]

\label{table:hstobs}
\end{longtable}
     \renewcommand{\thefootnote}{\arabic{footnote}}
     \renewcommand\tabcolsep{3pt}
}

\section{Results}

\subsection{Radio Analysis}

\begin{center}
\begin{figure}[t]%
\plotone{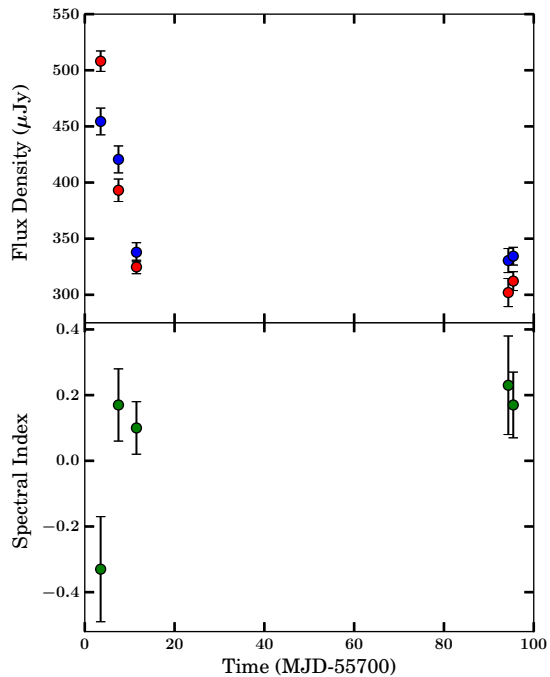}
  \caption{Radio frequency light curve of VLA J2130+12 (top) and corresponding spectral indices (bottom). In the top panel, red circles are the 5 GHz measurements and blue circles are the 7/7.45 GHz measurements.}%
    \label{fig:vlaspectra}%
\end{figure}
\end{center}

Table \ref{table:vladata} summarizes our fitted flux densities and derived spectral indices from each of our 4 epochs in May and August 2011 and Table \ref{table:vladataarc} summarizes the fitted flux densities for each available epoch in the archival VLA observations taken between 1989--2004. 
In the simultaneous 2011 data, the source is clearly variable (a factor of $\sim1.5$ variation in our measured flux densities), and the spectral index is primarily slightly inverted, albeit with an excursion to steep when it is at its brightest in the first epoch (see Figure \ref{fig:vlaspectra}). 
In the archival data, we see a factor of $\sim2$ variation in the measured flux densities at both 4.9 and 8.4 GHz.

Although the source is variable, we note that if we combine our 2011 VLA and archival VLA measurements with the flux density measurements reported in \citet{knapp1996} of $230\pm40$ $\mu$Jy at 8.4 GHz, an estimated 5$\sigma$ upper limit of $150$ $\mu$Jy at 4.9 GHz derived from archival data reported in \citet{machin1990}, and the flux density measurements at 1.6 GHz reported in K14 ($214\pm20$ $\mu$Jy), we also find that VLA J2130+12 is consistent with flat spectrum source. However, with this being said, the significant variability seen in the source
could easily cause large errors on non-simultaneous spectral indices. As such, we caution against drawing any conclusions from spectral indices calculated with non-simultaneous data separated by years.

The variability of VLA J2130+12 that we observed is consistent with the factor of $\sim6$ variation observed by K14 over the few-month timescale of their observations.  However, since there was not enough flux in the field to be able to self-calibrate, it is possible that uncorrected atmospheric phase wander (i.e., on timescales shorter than our phase calibration cycle time) may have led to some phase de-correlation, and hence introduced spurious variability to our measurements of VLA J2130+12.  We therefore searched for background check sources in the field, whose measured amplitudes could be used to assess the likelihood of any phase de-correlation.  While the previously known sources (AC211, M15 X-2, and source S1 of K14) all appeared to vary significantly, we found two sources located 2.5\,arcmin to the south and west of the pointing centre, with flux densities of 180 and 150\,$\mu$Jy\,bm$^{-1}$, respectively, that did not appear to vary significantly over the five epochs of observation.  This gives us confidence that the

\noindent observed variability of VLA J2130+12 (at a distance of 83\,arcsec from the pointing centre) is indeed real.

To determine the position of VLA J2130+12 on the $(L_R/L_X)$ plane, we derive a mean radio luminosity over all five 2011 observations. We find a luminosity of ${\rm(1.12\pm0.19)\times 10^{28} \, erg \, s^{-1}}$ at 5 GHz and ${\rm(1.61\pm0.26)\times 10^{28} \, erg \, s^{-1}}$ at 7 GHz. The errors in luminosity are quoted to the 1$\sigma$ confidence level and include the uncertainties in flux density and distance. Note that here and throughout this paper, we calculate radio luminosity by assuming a flat radio spectrum up to the observed frequency (i.e., $L_r=4\pi d^2 \nu S_{\nu}$ where $S_{\nu}$ is observed flux density, $\nu$ is the observing frequency, and $d$ is distance to the source).

\subsection{X-ray Analysis}

As the merged 102 ks Chandra/ACIS image shows no evidence for an X-ray counterpart at the location of VLA J2130+12 (see Figure \ref{fig:chandraimage}),  we estimate an upper limit on its brightness.
Before estimating this upper limit, we first corrected the astrometry in the Chandra images.
Due to heavy pile-up in ACIS frames from AC211 and M15 X-2, it is difficult to obtain accurate localization of these sources to match with radio coordinates. Thus, we corrected the absolute astrometry of the Chandra/ACIS image in two steps: (i) we corrected the astrometry of archival Chandra/HRC images using precise radio coordinates of AC211 (HRC detectors are not affected by pile-up), and (ii) we used coordinates of M15-X3 \citep{arnason2015} to match the ACIS image to HRC.

To estimate the upper limit we consider a 1\arcsec \,extraction radius, centered at the reported location of VLA J2130+12, which encircles $\sim$90\% of the energy\footnote{See Chandra Guide chap.4, fig. 4.23, at http://cxc.harvard.edu/proposer/ POG/html/chap4.html\#fg:hrma\_ee\_aspect\_hrc\_s\_point\_obs\_guide, for details.}. 
M15 contains two persistent X-ray binaries in its core (AC 211 and M15 X2), both with $L_X \geq 10^{36}$ erg/s. These sources produce large point spread function (PSF) wings in Chandra observations. To consider the effect of these sources, we chose four background extraction regions around these two sources at the same angular distance as VLA J2130+12 from them (see Figure~\ref{fig:chandraimage}). We estimate an upper limit on the count rate of $6.6\times 10^{-5}$ counts s$^{-1}$ (95\% confidence) in the 0.3--6 keV band.

Using PIMMS\footnote{http://asc.harvard.edu/toolkit/pimms.jsp}, we convert this count rate to flux by assuming a photon index of 1.5, typical for LMXBs in the hard state (e.g., \citealt{campana1998,campana2004}), and a column density $N_H=8.7\times10^{20}$ ${\rm cm^{-2}}$. This column density is calculated by assuming R$_V$=3.1, and using the reddening quoted for M15 (i.e., $\rm{E(B-V)=0.1}$)

\noindent from the Harris Catalog, 2010 Edition \citep{harris1996}, in combination with the correlation between $A_v$ and $N_H$ from \citealt{foight2015} and \citealt{bahramian2015} (i.e., $N_H=2.81\times10^{21} \, A_v$). As most of the reddening is occurring in the disk, this value is a valid approximation for VLA J2130+12, which is 2.2 kpc away in the direction of M15.
We find (95\% confidence) upper limits on flux in the 0.3--6 keV and 1--10 keV bands of $5.6\times10^{-16}$ and $ 6.4\times10^{-16}$ ${\rm erg\, cm^{-2}\, s^{-1}}$, respectively.
Assuming a distance of 2.2 kpc, yields an upper limit on luminosity of $3.2\times10^{29}$ ${\rm erg\, s^{-1}}$ in the 0.3--6 keV band and $3.7\times10^{29}$ ${\rm erg\, s^{-1}}$ in the 1--10 keV band.
With the derived radio and X-ray luminosities we have placed VLA J2130+12 on the $(L_R/L_X)$ plane (see Figure \ref{fig:LRLXplane}). Its position in the plane (in the radio-loud, X-ray quiet regime) provides further evidence supporting the presence of a BH in the system.

We have also calculated (95\% confidence) upper limits on flux and luminosity in the 0.3--6 keV band in all other available Chandra data (HRC and gratings; see Table \ref{table:chandraobs}) using the same method described above. In doing so, we find that in all Chandra observations the upper limits on the source has never been brighter than $4.6\times10^{-14}$ ${\rm erg\, cm^{-2}\, s^{-1}}$ (i.e., a few times $10^{31}$ ${\rm erg \, s^{-1}}$ at a distance of 2.2 kpc).

\subsection{Optical Analysis}

To determine whether the candidate counterpart identified in the ACS/WFC images is a cluster member, we plot the colour-magnitude diagram (CMD) of M15 using data from the ACS cluster survey (see Figure \ref{fig:cmd}). Note that we have made use of the quality flag\footnote{The quality flag ranges from 0 (perfect quality) to 2.5 (bad quality).} available in the ACS datasets on I and V band magnitudes for each source  \citep{sarajedini2007}, only plotting points with quality flag values below 0.5. The candidate optical counterpart is plotted in red. Note that magnitudes of sources in the ACS catalog are based on multiple measurements and the errors plotted are the rms residuals of the individual measurements.

The candidate optical counterpart appears to be redder than the main sequence in M15. Compared to the distribution of main sequence stars, this counterpart is a 3$\sigma$ outlier. While this strongly suggests that this object is not a member of the cluster, we note that there are other possibilities that could explain this observation (e.g., objects that appear redder than the main sequence due to unusual binary evolution paths, such as sub-subgiants; \citealt{mathieu2003}).

To determine how likely it is that the optical counterpart is in fact associated with VLA J2130$+$12 (rather then being a non-cluster member in this direction of the sky), we determined the radius at which the next 3$\sigma$ outlier from the main-sequence (plus horizontal branch) appears. This second outlier is $4.5\arcsec$ away from the radio coordinates; 237 sources, including the counterpart are closer than this 3$\sigma$ outlier. Given that our likely optical counterpart is only 0.08\arcsec away and the second 3$\sigma$ outlier is $4.5\arcsec$ from the radio coordinates of VLA J2130+12, we estimate the probability of finding a 3$\sigma$ outlier within a 0.1\arcsec radius of these coordinates purely by chance of $2\times(0.1^2)/(4.5^2) = $0.1\%. Thus, we conclude that this star is likely to be the counterpart of VLA J2130+12.

To constrain the nature of the identified optical counterpart we use \textit{Pysynphot} to fit the V and I magnitudes. We use Allard et al's Phoenix models, with parameter inputs from \citet{baraffe1998} and \citet{baraffe1996} so we can associate $T_{\rm eff}/\log g$ with a particular mass. Assuming that the source is at a distance of 2.2 kpc (and is hence associated with the radio source), we fit two different models. The first, a main sequence counterpart only model and the second, a two component, main sequence counterpart + power-law accretion disk model. Note that, given that the upper limits from the other (WFC3) filters are not very constraining and it is likely that neither of the best-fit models would predict a significant detection in any of these bands, we only make use of the V and I magnitudes in this analysis.

The spectrum of an accretion disk that is purely viscously heated will have three regimes: a Rayleigh-Jeans regime, at wavelengths longer than the Wien peak of the outermost annulus of the disk; a Wien spectrum at wavelengths shorter than the Wien peak of the innermost annulus of the disk; and a $\lambda^{-7/3}$ regime joining smoothly to the other two parts of the spectrum.
As such, in this analysis we fix the power-law index at $\Gamma_{\lambda}=-7/3$, corresponding to the ``flat'' part of an optically thick, geometrically thin accretion disk spectrum.

In both cases, we consider both a metal-poor ($[M/H]=$

\renewcommand{\thefootnote}{\alph{footnote}}
\renewcommand\tabcolsep{5pt}
\begin{longtable}{lcccr}

\caption{Pysynphot Phoenix Model Fits}  \\

\hline \hline \\[-2ex]
   \multicolumn{1}{l}{Fit ID$^a$} &
      \multicolumn{1}{l}{Metallicity$^b$} &
   \multicolumn{1}{c}{Counterpart Mass } &
     \multicolumn{1}{c}{${ \rm \chi^2/dof}$} &
          \multicolumn{1}{r}{$P_{\rm null}$$^e$} \\
     & \multicolumn{1}{l}{Assumption}& \multicolumn{1}{c}{$M_{\odot}$}&& \\[0.5ex] \hline
   \\[-1.8ex]
\endfirsthead

  \\[-1.8ex] \hline \\[-1.0ex] 
        \multicolumn{5}{p{0.95\columnwidth}}{NOTE. -- all errors are quoted to the 1$\sigma$ confidence level. For the solar and metal-poor fits, a distance of 2.2 kpc is assumed. For the M15 metallicity fit, the distance to M15 (10.4 kpc) is assumed. }\\
        \multicolumn{5}{p{0.95\columnwidth}}{$^a$ star-only = main sequence counterpart only model, and star+disk = main sequence counterpart + power-law accretion disk model. }\\
                \multicolumn{5}{p{0.95\columnwidth}}{$^b$ Metal-poor corresponds to $[M/H]=-1$, and M15 refers to the metallicity of the cluster of [M/H]$\approx -2$. See text for details. }\\
\multicolumn{5}{p{0.95\columnwidth}}{$^c$ Note that we find two minima for this model. This can occur in the low-mass regime where one colour (V-I) can correspond to multiple magnitudes ($m_v$). Given the data quality (i.e., zero dof) we can not confidently prefer one fit over the other. In addition, the known distance to VLA J2130+12 also does not allow us to prefer one fit over the other. }\\
                        \multicolumn{5}{p{0.95\columnwidth}}{$^d$ Note that this fit returned a model with zero flux in the accretion disk. }\\
                \multicolumn{5}{p{0.95\columnwidth}}{$^e$ Null hypothesis probability. }\\
\endlastfoot
star-only & solar&$0.150^{+0.025}_{-0.010}$&4.12/1&0.04 \\[0.1cm]
& metal-poor&$0.120\pm0.010$&1.36/1&0.24 \\[0.1cm]
&M15& $0.330\pm0.010$&7.30/1&0.007\\[0.25cm]
star+disk & solar$^c$&$0.145\pm0.010$&0.002/0& - \\
 & &$0.185\pm0.010$&1.75/0&- \\[0.1cm]
& metal-poor&$0.115\pm0.010$&0.006/0&- \\[0.1cm]
& M15$^d$&$0.330\pm0.010$&7.30/0&- \\[-2.5ex]

\label{table:synphotfits}
\end{longtable}
     \renewcommand{\thefootnote}{\arabic{footnote}}
     \renewcommand\tabcolsep{3pt}

\noindent $-1.0$) counterpart and a solar metallicity ([M/H]=0.0) counterpart. The best-fit results are presented in Table \ref{table:synphotfits}. Overall, the fit results suggest a low-mass companion, between $\sim0.1-0.2 M_{\odot}$. The metal-poor star only model produces an acceptable fit ($P_{\rm null} > 0.1$), while the solar abundance star only model produces a marginally unacceptable fit. Since the addition of a disk eliminates all degrees of freedom, we can not calculate a null hypothesis probability. Instead, we can consider acceptable fits to occur when $\chi^2=0$. For comparison with $P_{\rm null} > 0.1$, we list all models within $\chi^2 < 2.71$, the $\chi^2$ that corresponds to 90\% confidence interval for 1 degree of freedom. Similarly, we can compare the $\chi^2$ of the star and the star plus disk models. For the solar abundance model, the star plus disk model is consistent with no disk within the 95.7\% confidence interval; for the metal-poor abundance model, the star plus disk model is consistent with no disk within the 75.6\% confidence interval. Thus the model fits are suggestive, but no conclusive (i.e., not $3\sigma$) that we require an accretion disk to explain the counterpart's optical colours.

We also test the assertion that the identified counterpart is a member of M15 by fitting an additional model involving a main-sequence star at M15's distance ($10.3\pm0.4$ kpc; \citealt{vandenbosch2006}) and metallicity. Here we used the value quoted in the Harris catalog \citep{harris1996} of [Fe/H]=-2.37 for the metallicity of M15.  However, as Pysynphot's synthetic spectra use [M/H] as inputs, rather then [Fe/H], we took the reasonable assumption (also used in \citealt{baraffe1998}) that M15's [Fe/H] translates to an $[M/H]\approx-2.0$.
As the $\chi^2$ values of this model are much poorer than those found for the star-only fits (assuming either solar metallicity or a metal poor counterpart), the proposed optical counterpart is almost certainly not a cluster member (as argued previously; e.g. see green points in Figure \ref{fig:cmd}), and is therefore likely to be associated with the radio source. In addition, since accretion disks are bluer than the main sequence, adding an accretion disk to a main sequence model does not improve fits compared to the star-only model at the distance and abundance of M15.

Overall, given the previously discussed astrometric measurements, the significant radio variability observed on timescales of months, the observed flat to slightly inverted radio spectrum of the source, and the upper limit on the X-ray luminosity, it is likely that VLA J2130+12 a foreground LMXB containing a BH. If this is the case, to our knowledge VLA J2130+12 is the most 
radio-loud quiescent BH source known to date.

While the radio to X-ray flux ratio for VLA J2130+12 is abnormally high, even for a quiescent BH (see Figure \ref{fig:LRLXplane}), we note that high radio/X-ray flux ratios have been observed to 
occur in near face-on systems (e.g., MAXI J1836-194; \citealt{r14b}).     
 In addition, as some early works on advection dominated accretion flows (ADAFs; \citealt{ny94}) predict that the emission from the hot inner flow can start to fall out of the X-ray band at very low Eddington fraction, it is possible that most of the emission in VLA J2130+12 is at far-UV rather than X-ray frequencies, implying that the observed high radio/X-ray flux ratio could be explained in this case without invoking an enormous beaming factor. The large radio luminosity may also indicate a higher than average stellar mass BH in the system, if one applies the fundamental plane of BH activity (e.g., \citealt{merloni2003,falke2004}). Implications of this possibility are discussed further in Section 4.2.

\section{Discussion}
First, we discuss the nature of VLA J2130+12, offering alternative explanations that could possibly describe this source. Second, we discuss the binary properties of VLA J2130+12. Third, we discuss (i) how the discovery of this source is suggestive of a larger population of field BHXBs than has been typically assumed, and (ii) the implication that this larger number has on population synthesis modelling of these types of systems.

\subsection{The Nature of VLA J2130+12 - Alternative Explanations}
\label{subsection:S2type}
While the X-ray, radio, and optical properties of VLA J2130+12 indicate the presence of a stellar-mass BH (e.g., high radio/X-ray flux ratios, its specific position on the $L_R$-$L_X$ plane, photometry consistent with a main sequence companion star plus potential accretion disk at the foreground distance of the radio source), here we discuss alternative explanations that could possibly produce the properties of this source. A variety of Galactic sources emit, in the flux/luminosity range typical of BHXBs, at X-ray and/or radio wavelengths. However, many of these objects can be ruled out using a combination of X-ray and radio flux ratios, characteristic shape of the radio spectrum, and observations at other (optical/infrared) wavelengths (see e.g., \citealt{seaquist1993,maccarone2012}).

\subsubsection{Accreting Neutron Star Low-Mass X-ray Binaries (NS-LMXBs)}
We can rule out NS LMXBs based on a combination of X-ray/radio flux ratio and spectral index. Actively accreting NSs tend to be far less radio luminous (have much lower radio luminosities at a given X-ray luminosity) when compared to BH LMXBs (see e.g., \citealt{migliari2006} and observe the measurements for NSs and transitional milli-second pulsars (tMSPs) in Figure \ref{fig:LRLXplane}). Moreover, as tMSPs go into pulsar mode when they are faint, at a very low accretion rates onto a NS, the tMSP scenario is not consistent with our observations.
Therefore, we expect that it is exceptionally unlikely for a NS to be as radio luminous in quiescence as VLA J2130+12.

\subsubsection{Cataclysmic Variables (CVs)}
\begin{center}
\begin{figure*}%

\plotone{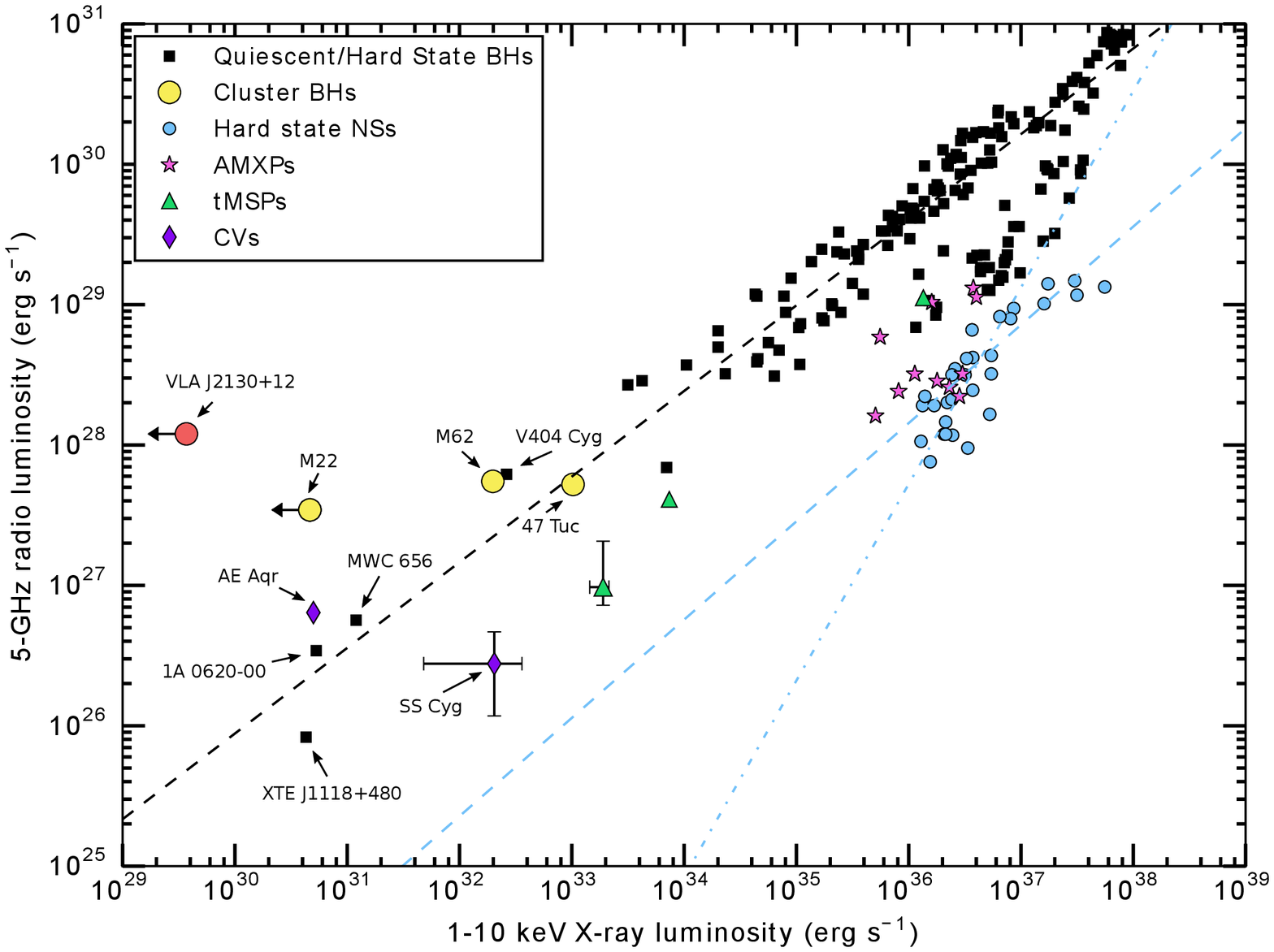}
  \caption{Radio/X-ray correlation for stellar-mass BHs in the hard and quiescent states. VLA J2130+12 is represented by the red circle. The black squares represent field BHs from the literature \citep{millerjones2011,gallo2012,ratti2012,co13,gallo2014,munar14,dzib2015}, and the yellow circles represent the BHCs in M22, M62, and 47 Tuc \citep{strader2012,chomiuk2013,millerjones2015b}. The blue circles represent NS systems in the hard state (\citealt{rutledge1998,moore2000,migliari2003,migliari2006,migliari2010,tudose2009,millerjones2010,migliari2011,tet16}). Green triangles and pink stars show the binary transitional milli-second pulsar \citep{hill2011,papitto2013,deller2015} and accreting milli-second pulsar \citep{gaensler1999,rupen2002b,pooley2004,fender2004c,rupen2005a,migliari2011} systems, respectively. 
  The purple diamonds show the CVs, AE Aqr \citep{eracleous1991,abadasimon1993} and SS Cyg (Russell et al. 2016, in prep.), during bright flare/outburst periods in each source. These points are meant to represent the most radio bright periods observed in these types of systems. The vast majority of CVs lie below these points in the $L_R/L_X$ plane \citep{fuerst1986,kording2008,kording2011,byckling2010}.
  The dotted black line shows the best fit relation for BHs \citep{gallo6}, and the blue dashed and dashed-dotted lines show the two suggested correlations for NS systems \citep{migliari2006}. 
}%
    \label{fig:LRLXplane}%
\end{figure*}
\end{center}

\begin{center}
\begin{figure}[h!]%
\plotone{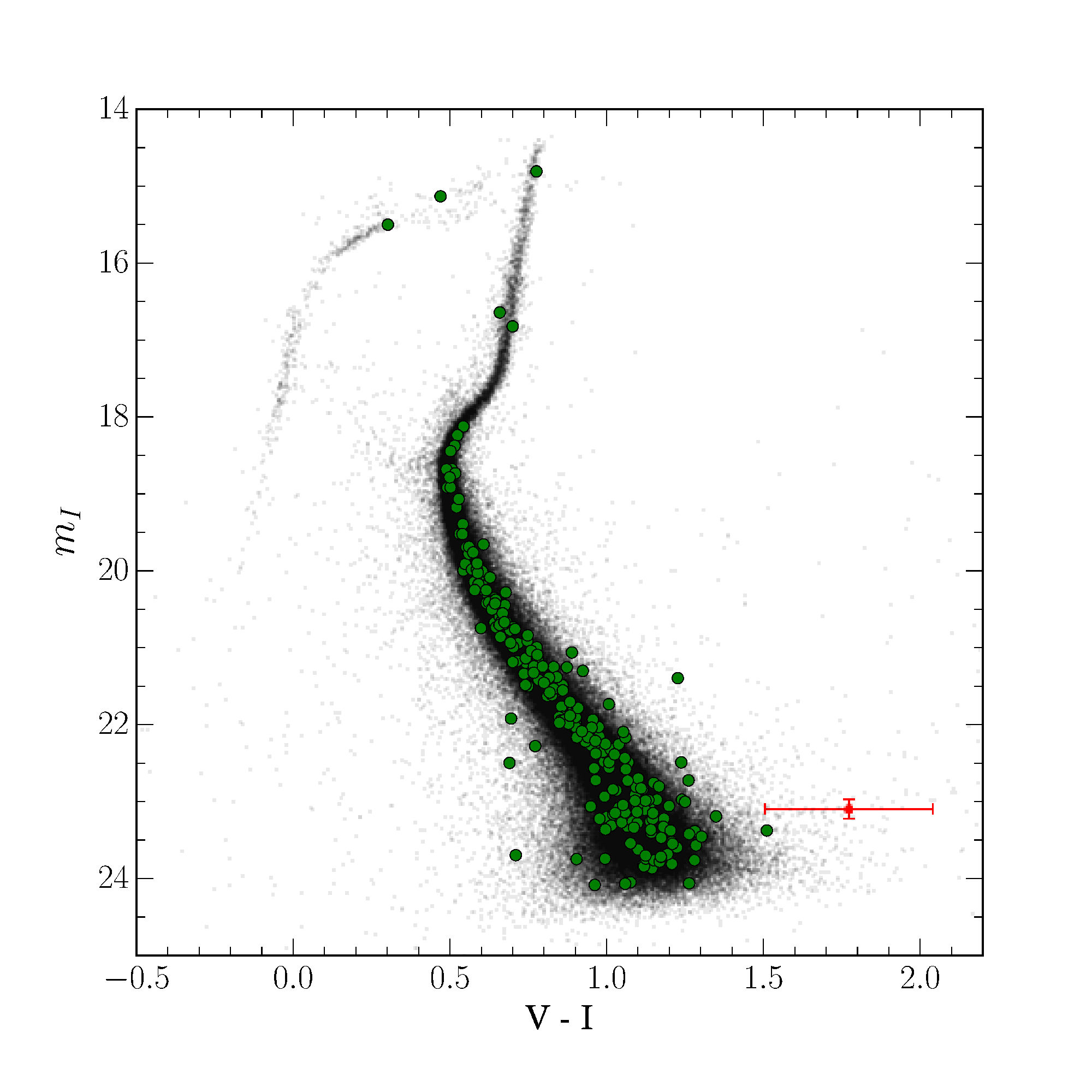}
  \caption{Colour-magnitude diagram of M15 made using data from the ACS cluster survey. The data shown has been filtered by the quality flag for I and V band magnitudes available in ACS datasets. Only points with quality flag values below 0.5 are plotted here. Green points indicate sources within a 4.5\arcsec radius around the radio coordinates of VLA J2130+12. The identified optical counterpart of VLA J2130+12 is plotted in red. All uncertainties are directly quoted from ACS database. The position of VLA J2130+12 strongly indicates that it is not a cluster member and therefore is most likely associated with the radio source (see text for details).}%
    \label{fig:cmd}%
\end{figure}
\end{center}

We can also rule out CVs based on their radio and X-ray fluxes. In Figure \ref{fig:LRLXplane} we plot simultaneous radio/X-ray data for SS Cyg and AE Aqr during brief, bright flares/outbursts in each source. The vast majority of CV radio measurements are below the two points plotted (see e.g., \citealt{fuerst1986,kording2008,kording2011,byckling2010}). While VLA J2130+12 is variable,
at its faintest measured radio luminosity it is (i) nearly three orders of magnitude more luminous then the persistent quiescent emission detected from any known CV (e.g., \citealt{mason2007}), and (ii) more than an order of magnitude more luminous then the brightest radio emission detected during the peak of flare/outburst events observed in nova-like CVs (e.g., \citealt{coppejans2015}). As such, it is unlikely that a quiescent CV could reach radio luminosities as high as VLA J2130+12.

\subsubsection{Magnetars}
Magnetars can be radio-bright, variable, have a flat spectrum, and show X-ray luminosities $<10^{31} {\rm erg \, s^{-1}}$ (see \citealt{olausen2014}), like VLA J2130+12.
However, from both a theoretical and observational standpoint, magnetars are young objects; their observed positions with respect to the Galactic plane indicate an age $<10^4$ years \citep{olausen2014}. 
A lower limit on the age of VLA J2130+12 can be inferred by integrating the trajectory of the source backwards in time from the observed Galactic position and proper motion, and computing the time at which the source crosses the Galactic plane (assumed as birthplace; see Section 4.2). To do so, we assume the distance to VLA J2130+12 to be uniformly distributed in the range $[1.9, 2.7]$ kpc (K14) and the proper motion to be distributed as a normal distribution in both $\alpha$ and $\delta$ directions, taking as standard deviation the errors on the measurement. Since the line of sight velocity is not known, we model its value as being uniformly distributed in the range $[-300, 300]$ km/s. We integrate $5\times 10^4$ orbits backwards in time for $5$ Gyr 
using the Python package for galactic dynamics {\tt{\small galpy}} \citep{bovy2015}, with a Galactic potential from \citealt{irrgang2013}. By recording the times at which the binary crosses the Galactic plane, which we define as $z< 0.1$ kpc, we find a lower limit for the crossing time of $\approx 7$ Myr. If instead, we only integrate $5\times 10^4$ orbits backwards for the magnetar age limit of $10^4$ years, we find the source would of have to have been born between 0.88 and 1.22 kpc below the Galactic plane. Since this is a factor of 4 more distant from the Galactic plane than any known magnetar or magnetar candidate \citep{olausen2014}, this clearly rules out the magnetar possibility.

\subsubsection{Planetary Nebula (PNe)}
PNe emit optically-thin thermal emission at radio wavelengths (and hence show a flat spectral index) and have low X-ray luminosities (i.e., $\sim10^{30} \, {\rm erg \, s^{-1}}$; 
\citealt{montez2010}), both consistent with VLA J2130+12. In fact, there is a known PNe in M15 (K648; \citealt{odell1964}), which is a $\sim$4 mJy 5GHz radio source \citep{birkinshaw1981} and a weak X-ray source ($10^{31}-10^{32} {\rm erg \, s^{-1}}$, \citealt{hannikainen2005}). However, no extended emission is observed around VLA J2130+12 in the wide array of archival HST images we have acquired. In addition, (i) the observed optical colours are not consistent with the presence of a hot white dwarf, expected to be within the PNe itself, and (ii) the VLBI flux density measurements from K14 indicate brightness temperatures of $>4\times10^6$ K, inconsistent with optically-thin thermal emission ($<10^4$ K) from a PNe.
Lastly, given PNe typically lack any radio variability, and display significantly extended radio emission (e.g, 3\arcsec for the more distant K648), 
neither of which apply to VLA J2130+12, this possibility is unlikely.

\subsubsection{Millisecond Pulsars (MSPs)}
While MSPs can display radio/X-ray flux ratios in the range of interest (see e.g., \citealt{possenti2002,maccarone2012}), these sources typically display steep radio spectra, inconsistent with the primarily flat spectrum we observe for VLA J2130+12. See for example \citealt{bates2013}, who find a spectral index distribution with a mean of $-1.4$ and a standard deviation of $0.96$ for these systems, or the rotation powered pulsars in the ATNF Pulsar Catalog \citep{manchester2005}. Pulsars in the field can occasionally drive a wind that interacts with ambient material, called a pulsar wind nebula. However, while pulsar wind nebulae show a flat radio spectrum, they generally have X-ray luminosities of $L_X>10^{34} \, {\rm erg \, s^{-1}}$, large diameters, short lifetimes, and dense surroundings \citep{gaelsler2006}, all inconsistent with our observations of VLA J2130+12. 
In addition, we note that the pulsar population in the field of M15 has been extremely well studied\footnote{See http://www.naic.edu/$\sim$pfreire/GCpsr.html for a list of the M15 discovered pulsars and relevant references.}, including a thorough acceleration search by \citealt{anderson1993}.
For VLA J2130+12 to be a millisecond radio pulsar, it would have to show several unusual features (a flat radio spectrum, a very small X-ray luminosity) and would have to be beamed away from Earth, which is unusual for millisecond pulsars considering their large beam sizes (e.g. \citealt{lyne1988,heinke2005}).

\subsubsection{Ultracool Dwarfs}
While coronally active ultracool dwarfs (i.e., later than M6) can have radio/X-ray flux ratios in the range of interest and display flat radio spectra, these objects cannot be as radio luminous as VLA J2130+12 ($L_R\lesssim10^{10}-10^{14} \, {\rm erg \, s^{-1}}$; see e.g., \citealt{berger2010,maccarone2012}). In addition, we note that these objects have only been observed to undergo moderate variability, therefore it is likely that we are not just observing a flare in the source. As such, this possibility is ruled out.

\subsection{System Parameters of VLA J2130+12}

Binary counterpart masses of the Galactic BH-LMXB population range from $\sim$0.18--3.0$M_{\odot}$ (e.g., see \citealt{tetarenkob2015}). With an estimated counterpart mass of $\sim0.1-0.2 \, M_{\odot}$, VLA J2130+12 is one of the smallest BHXB companions known. Assuming a Roche lobe filling star (e.g., see \citealt{frank2}), we estimate an orbital period for the system of $\sim1-2$ hours, making VLA J2130+12 one of only 7 known BH-LMXBs that belong to the short period (< 5 hours) regime (see \citealt{port2004,tetarenkob2015}).

The Galactic coordinates of VLA J2130+12 $(l, b) = (65.009841, -27.334497)$, together with its distance, indicate a Galactic position at $(R,z)\approx (7.4, -1.0) $ kpc,
where $R$ is the distance from the Galactic centre projected on to the Galactic plane, and $z$ is the distance from the Galactic plane. This $z$ places this system in the subset of BHXBs at large offset from the Galactic plane ($z\gtrsim 1$ kpc; see Table 5 in \citealt{repetto2012}, and Table 2 in \citealt{repetto2015}). Given the set of orbital parameters and its current position, we can investigate whether the measured space velocity of VLA J2130+12 is consistent with an object originally in the disk that received a kick up into the halo.

Since BHXBs are thought to be formed in the plane of the Galaxy where most of the massive stars reside, a large distance from the plane is suggestive for a natal kick (NK) at birth, which is the velocity potentially acquired by a BH at formation. Assuming the binary received a velocity at birth perpendicular to the plane $V_\perp$ as a result of the NK, and that it travels from its birth place to the observed $z$, we can estimate a lower limit for its peculiar velocity at birth as $v_{\rm pec, min}=V_{\perp}$. The $V_{\perp}$ is obtained by conservation of energy along the trajectory in the Galactic potential (see \citealt{repetto2012,repetto2015}). Accounting for the uncertainty in the distance to VLA J2130+12, we find a velocity $V_{\rm pec, min}$ in the range $\approx 59-75$ km/s. To convert such velocity into a NK, 
we trace the binary evolution of the source backwards in time, from the observed properties (BH mass, companion mass, and orbital period) until the BH formation (see e.g., \citealt{repetto2015}). Here we take a standard BH mass of $10~M_\odot$ and a companion mass of $0.2~M_\odot$, which give an orbital period of $P_{\rm orb} \approx 2$ hours.
The orbital properties at BH formation, together with the lower limit on the peculiar velocity $v_{\rm pec, min}$, deliver a minimum NK of $80-100$ km/s. Such a value for the NK would make this system similar to XTE J1118+480, a BHXB whose formation requires a NK of at least $\approx 80$ km/s \citep{fragos2009}. Hence, we conclude that VLA J2130+12 is consistent with having been formed in the disk and later kicked up into the halo of the Galaxy. This result potentially enlarges the sample of BHXBs that require a NK at formation.

There are 4 other known Galactic BH systems with a short period and high Galactic latitude like VLA J2130+12. It has been suggested that this subset of systems was also kicked out of the Galactic plane into the halo (e.g., see \citealt{ku13}). We note that both this observation and our result are consistent with what is expected from theory, as wider orbits should be more easily disrupted when the BH forms and receives a NK, due to the lower binding energy of the system. Whereas tighter orbits can survive higher NKs (e.g., see \citealt{brandt1995}).

Given that one possible explanation for the anomalously high radio/X-ray flux ratio of VLA J2130+12 (compared to other quiescent BHs) is that the system may contain a BH of $>10 \, {\rm M_{\odot}}$, we have also considered what effect a larger mass would have on the computed NK. From a theoretical standpoint, the physics of the NK is not yet agreed upon (see e.g., \citealt{fryer2006} for discussion). 
If for example, the BH is formed via fallback on to the proto neutron star and the NK is caused by asymmetric mass ejection in the supernova, then we would expect the NK to decrease with increasing BH mass \citep{fryer2012}, as higher-mass BHs eject less mass at formation.
However, (i) there are other equally viable models in which this is not the case (e.g., neutrino-driven NKs; \citealt{janka2013,belczynski2016}) and (ii) observationally, there is no clear correlation seen between NK and BH mass (see Figure 12 in \citealt{repetto2015}). Specifically, for our simulations, the only observational constraint we have is $V_{\rm pec,min}$ (which has been computed assuming the binary was born in the Galactic plane, as described above). The peculiar velocity $V_{\rm pec}$ that a binary acquires as a result of the BH formation
is a combination of the NK and of the recoil that the binary receives as a consequence of the mass ejection in the supernova. The larger the total mass of the binary,
the lower $V_{\rm pec}$ is. 
Thus a binary containing a larger-mass BH would require a larger NK to have the same peculiar velocity of a binary containing a lower-mass BH. This implies that 
our calculated lower limit on the NK is a conservative estimate, and would be valid even if the system did contain a $>10 \, {\rm M_{\odot}}$ BH.

On the other hand, because the companion star is inferred to be of very low mass, we cannot exclude that the system was born longer than $10$ Gyr ago in the thick disk. The thick disk current scale height is $\approx1$ kpc \citep{juric2008}. Evolutionary sequences of BHXBs that host low mass stars as companions ($M_2\approx 0.1-0.4~M_\odot$), indicate a typical transferred mass to the BH between $0.7-1.5~M_\odot$ (see Table 3 in \citealt{fragos2015}).
We can hence not exclude that the companion star was $\approx 0.8~M_\odot$ on the zero-age main sequence, which would make a thick disk origin for VLA J2130+12 plausible.
However, the scale height of the thick disk could
be much smaller in the past than currently (due to subsequent heating). Taking an initial distance from the Galactic plane of $z=0.3$ kpc,
the lower limit on the NK is only slightly affected: $74-92$ km/s. On the other hand, we can also not exclude that the companion star
was initially more massive than $\approx 1~M_\odot$,  thus we find the hypothesis of a birth in the thin disk as the most likely one.
Other possible origins, other than a birth in the Galactic plane, include (i) the formation and consequent ejection of the binary from a globular cluster, and (ii) the formation of the binary in the halo. However, since the metallicity of the companion star is currently not strongly constrained,
it is not yet possible to assert the likelihood of these two scenarios.

Lastly, with the determined set of orbital parameters we can postulate on the X-ray outburst properties of VLA J2130+12.
Binary evolution indicates that short period systems like VLA J2130+12 should not only exist but also dominate the total BHXB number counts (e.g., \citealt{knev14}). 
Short-period XRB (BH and NS) systems are expected to have low peak outburst luminosities because their small disks only hold a small amount of mass (see \citealt{king2000,kingrit8,wu2010}). In addition, it is thought that BH-LMXBs are thought to be increasingly inefficient to X-ray luminosity with decreasing mass transfer rates (e.g. \citealt{knev14}), making outbursts in short period BH-LMXBs peak at even fainter luminosities.
It has even been suggested that the temperature of the radiation during these outbursts could drop out of the X-ray band into the far-UV, so that even outbursts from nearby sources would not be detected by current surveys \citep{shab98,wu2010,maccarone2013,knev14}.

As such, given the estimated $P_{\rm orb}\sim1-2$ hours, we would expect VLA J2130+12 (and other systems like it) to have peak X-ray outburst luminosities that are well below (see Figure 2 in \citealt{knev14}) the typical detection threshold of the current all-sky and scanning instruments ($L_{\rm bol}<$ a few times $10^{36} \, {\rm erg s^{-1}}$; \citealt{tetarenkob2015}).
Thus, it stands to reason that, as the all-sky and scanning instruments are our most effective resource for discovering new transients in the Galaxy, there is likely a huge population of extremely low accretion rate BHXBs within our Galaxy, yet to be discovered (e.g., see \citealt{maccarone2015}).

\subsection{The Total Population of Field BH-LMXBs in the Galaxy}

To estimate how many quiescent field BH-LMXBs exist in the Galaxy, given that we find one such system in a random line of sight, we consider a simple population model for the density of BHXBs in the Galaxy. Given that VLA J2130+12 is located $\sim1$ kpc below the plane (see Section 4.2), and the fact that the maximum detected distance from the plane of the Galaxy for BH-LMXBs is 1.5 kpc \citep{repetto2012}, we start by modelling the Galaxy as a disk of radius 10 kpc and thickness 3 kpc (1.5 kpc above and below the plane). This yields a total volume of $9.4\times10^{11} \, {\rm pc^3}$. Next, we uniformly distribute $N$ quiescent BHXBs like VLA J2130+12 within this volume. As BHs are more concentrated towards the plane than towards the halo (e.g., see \citealt{jonker2004,repetto2012}), this choice will give us a conservative lower limit on the number of BH LMXBs in the Galaxy.

Next, we determine the volume of space that has been searched for systems like VLA J2130+12. VLA J2130+12 was primarily identified as a candidate BHXB based on its parallax measurements and measured radio properties. All the sources for the original project (K14) were originally chosen based on their properties measured in an Arecibo primary beam centred on the globular cluster M15.
Given that the Arecibo primary beam is 2\arcmin \,in radius, the radius of the surveyed cone will be 1.3 pc at d=2.2 kpc. Considering the position of VLA J2130+12 at $b=-27\deg$ and, furthermore, assuming it to be at the maximum distance of $1.5\,$kpc below the plane would yield a maximum distance of $3.3\,$kpc from Earth. Accordingly, this would yield a maximum radius of $1.9\,$pc and in-turn a total searched volume of $1.2\times10^4\,$pc$^3$.

We note that the volume considered here should actually be the total volume of space wherein observed serendipitous radio sources have measured parallaxes. In other words, we need to account for the ``number of trials'' effect, whereby the predicted number of quiescent BH-LMXBs would be scaled down by a factor equal to the number of observations like VLA J2130+12 that have been done (e.g., wide field VLBI parallax). The field of view of VLBI observations is traditionally limited by both bandwidth and time-average smearing,
effectively restricting the field of view to just a few arcseconds squared. As such, only a tiny portion of the primary beam of typical VLBI observations is actually analyzed. Additionally, as the space density of bright, highly compact radio sources is low, the entire field of view (even that limited by bandwidth/time-averaged smearing) is not usually imaged because of the low probability of detecting sources other than the target source.
We are not aware of similar searches for VLA J2130+12-like systems in other parallax observations, making the ``number of trials'' penalty very small, especially given the low flux density of VLA J2130+12.

Although this penalty is small, we can still provide a quantitative estimate based on a few assumptions. The observations of M15 that measured the parallax to VLA J2130+12 (K14) employed an interferometric technique where the position of a weaker potential in-beam calibrator source (M15 S1 and VLA J2130+12) can be measured against a brighter primary calibration source, with the potential for using the in-beam calibrator to transfer more accurate calibration solutions to other in-beam targets. Based on a literature review for papers where VLBI parallax measurements were determined using an in-beam calibrator, we find 24 other similar measurements, the majority of which have been used to measure parallaxes to pulsars \citep{fomalont1999,brisken2003,chatterjee2005,chatterjee2009,ng2007,middelberg2011,deller2012,deller2013,ransom2014,liu2016a}. As these measurements represent additional opportunities to detect the parallax of a VLA J2130+12 like object, they could be considered as potential trials. However, all of the reported in-beam calibrators were significantly brighter (4--86 mJy) than VLA J2130+12 ($\sim 0.1$--0.5 mJy). Given our conservative assumption of a uniform volume density of VLA J2130+12 like objects in the Galaxy, we should down-weight the number of trials from brighter sources by $(f_{\nu,\rm{in-beam}}/f_{\nu,\rm{VLA J2130+12}})^{-1.5}$. In that case, the number of trials penalty is only an additional $\sim0.2$ trials. In addition, we note that the PSR$\pi$ parallax project (a large VLBA program) have reported\footnote{\url{https://safe.nrao.edu/vlba/psrpi/}} 111 additional in-beam calibrator sources that they used to measure parallaxes. Although they do not provide flux densities of individual sources, they note that their median in-beam calibrator source is 9.2 mJy. We have measured the flux density function of secure 1--20 mJy FIRST sources (\citealt{helfand2015}; $dN/df_{\nu}\propto f_{\nu}^{-1.7}$) to estimate the expected distribution of  the flux densities of in-beam calibrators. We found that the minimum flux density is likely $\sim 3.2$ mJy, and using the same down-weighting we estimate an additional penalty of $\sim0.9$ trials.

These estimates imply a total number of trials penalty of $\sim2.1$ (including the VLA J2130+12 trial). We conservatively increase this penalty to 4 to account for any issues with our down-weighting scheme and unidentified analog observations. 
Considering this penalty, the fact that there are about 78 million volumes in the Galaxy similar to this one, and that the 3$\sigma$ confidence interval \citep{geh86} on the detection of 1 source is $0.0014-8.9$,
we estimate $2.6\times10^4$--$1.7\times10^8$ objects (i.e., quiescently accreting BHXBs) like VLA J2130+12 in the Galaxy. The lower bound of this conservative estimate is consistent with the upper limit of the theoretical estimates for the number of BH-LMXBs in the Galaxy computed from population synthesis codes ($\sim10^2-10^4$; \citealt{romani1992,romani1994,portegieszwart1997,kalogera1998,pfahl2003,yungelson2006,kiel2006}). This implies that the population synthesis estimates for the numbers of BHXBs undergoing mass transfer in the Galaxy are likely underestimated.

We postulate that a likely reason for this discrepancy stems from the fact that these population synthesis estimates are generally calibrated from the bright XRB population, ignoring what is thought to be the most numerous class of XRBs, the so called very faint X-ray transients (VFXTs), which are thought to be missing from the observed population because they undergo outbursts that are too faint to be detected by typical transient surveys.

\section{Summary}
The known radio source VLA J2130+12 \citep{knapp1996} is not associated with the Galactic GC M15. Based on proper motion and parallax measurements, along with observed variability at radio wavelengths (on timescales of months), K14 suggested that VLA J2130+12 is a probable field LMXB. We argue that this system is likely to contain a stellar-mass BH accreting from a low-mass ($\sim0.1-0.2\,M_{\odot}$) counterpart, making it the first known field BH-LMXB candidate system identified in quiescence. This argument is supported by its X-ray and radio luminosities, radio spectra, and the identification of the likely optical counterpart.

First, we observe a flat to slightly inverted radio spectrum for the source, consistent with BHXBs in the quiescent and hard accretion states. In addition, we find a radio luminosity of ${\rm1.2\times 10^{28} \, erg \, s^{-1}}$ at 5 GHz and an upper limit on X-ray luminosity of $3.7\times10^{29}$ ${\rm erg\, s^{-1}}$ in the 1--10 keV band. These luminosities make VLA J2130+12 the most radio-loud quiescent BH candidate currently known, and puts VLA J2130+12 at a position in the $(L_R/L_X)$ plane that strongly rules out accreting neutron star and white dwarf systems. In addition, we note that all other alternate scenarios we have considered to describe VLA J2130+12 have been ruled out (see Section \ref{subsection:S2type} for details). 

Secondly, given the derived orbital parameters of VLA J2130+12 (companion mass and orbital period), we find the current position of the system in the Galaxy to be consistent with it being born in the plane and receiving a large NK at birth ($\gtrsim80-100$ km/s), effectively launching it into the halo. We note that this finding adds VLA J2130+12 to the growing sample of (candidate and confirmed) BHXBs that require a NK at formation.

Third, we argue that the short orbital period ($\sim 1-2$ hours) of VLA J2130+12 would likely give rise to outbursts which radiate outside of the X-ray band, primarily in the UV. Making outbursts from VLA J2130+12, and other sources like it located within a few kpc of Earth, undetectable by current all-sky and scanning X-ray instruments, which are largely our most effective resource for discovering new transients in the Galaxy to date.

Lastly, using the fact that we observe one quiescent BH-LMXB in a random line of sight, we estimate a total number of Galactic quiescent field BH-LMXBs between $2.6\times10^4-1.7\times10^8$, suggesting that a large population of extremely low accretion rate BHXBs exist within the Galaxy. Considering, (i) the relatively small area over which deep radio/X-ray studies of radio-loud/X-ray quiet sources have been performed, and (ii) the fact that the lower limit of this conservative estimate is consistent with the high end of the numbers computed from population synthesis codes, it becomes clear that to identify more of the Galactic BHXB population, and further our understanding of the formation and evolution of these types of systems, large-scale radio or X-ray surveys are needed. 
Possible methods that could be used to accomplish this goal in the near-term include (i) taking multiple VLBA position measurements of bright, Galactic, flat-spectrum radio sources to look for high proper motions that would indicate nearby objects, accompanied by follow-up observations at other wavelengths upon detection, or (ii) performing repeated, deep, wide X-ray surveys of the bulge region (where most of the known BH-LMXBs are located) to look for faint ($L_{\rm X, peak}<10^{36} \, {\rm erg \, s^{-1}}$) X-ray outbursts of BH LMXBs that would normally go undetected by typical all-sky surveys, followed by observations at other wavelengths if one is detected. While these surveys should be started with current instruments, the next generation of both X-ray telescopes with large sensitivity over wide field-of-views and radio arrays capable of more sensitive astrometry may be needed to identify and confirm the nature of large numbers of this population (e.g., \citealt{maccarone2005,fender2013}).

\section{Acknowledgments}
The authors would like to thank Tom Russell for providing the radio data for SS Cyg and Chris Done for useful discussions. BET, GRS, and COH acknowledge support by NSERC Discovery Grants. JS acknowledges support from NSF grant AST-1308124. JCAMJ is the recipient of an Australian Research Council Future Fellowship (FT140101082). WV acknowledges financial support from the Swedish Research Council. The National Radio Astronomy Observatory is a facility of the National Science Foundation operated under cooperative agreement by Associated Universities, Inc.
The search for an optical counterpart was based on observations made with the NASA/ESA Hubble Space Telescope, and obtained from the Hubble Legacy Archive, which is a collaboration between the Space Telescope Science Institute (STScI/NASA), the Space Telescope European Coordinating Facility (ST-ECF/ESA) and the Canadian Astronomy Data Centre (CADC/NRC/CSA). The scientific results reported in this article are in part based on observations made by the Chandra X-ray Observatory.  This research has made use of software provided by the Chandra X-ray Center (CXC) in the application package CIAO and the pysynphot package developed as part of STSDAS by the Space Telescope Science Institute (STScI). This publication makes use of data products from the Two Micron All Sky Survey, which is a joint project of the University of Massachusetts and the Infrared Processing and Analysis Center/California Institute of Technology, funded by the National Aeronautics and Space Administration and the National Science Foundation.
This work has also made use of NASA's Astrophysics Data System (ADS).

\bibliographystyle{apj.bst}

\bibliography{fullreflist.bib}

\end{document}